\documentclass[12pt,a4paper]{JHEP3}

\usepackage{picinpar}
\usepackage{epsfig}
\usepackage{amsmath}
\usepackage{amssymb}
\usepackage{cite}
\usepackage{graphicx}

\newcommand{\beqs}{\begin{equation*}}

\newcommand{\beq}{\begin{equation}}

\newcommand{\eeqs}{\end{equation*}}

\newcommand{\eeq}{\end{equation}}

\newcommand{\beqas}{\begin{eqnarray*}}

\newcommand{\beqa}{\begin{eqnarray}}

\newcommand{\eeqas}{\end{eqnarray*}}

\newcommand{\eeqa}{\end{eqnarray}}

\newcommand{\eq}[2]{\begin{equation} #1 \label{#2} \end{equation}}

\newcommand{\eps}{\varepsilon}

\newcommand{\ga}{\gamma}

\newcommand{\de}{\delta}

\newcommand{\blist}{\begin{itemize}}

\newcommand{\elist}{\end{itemize}}

\providecommand{\href}[2]{#2}

\DeclareFontFamily{OT1}{rsfs}{}

\DeclareFontShape{OT1}{rsfs}{m}{n}{ <-7> rsfs5 <7-10> rsfs7 <10->rsfs10}{} 

\DeclareMathAlphabet{\mycal}{OT1}{rsfs}{m}{n}

\DeclareMathOperator{\extdm}{d}

\newcommand{\extd}{\extdm \!}
\newcommand{\Tr}{{\rm Tr}\;}

\title{Non-AdS holography in 3-dimensional higher spin gravity --- General recipe and example}

\author{H.~Afshar, M.~Gary, D.~Grumiller, R.~Rashkov\footnote{On leave from the Department of Physics, Sofia University.} \ and M.~Riegler\\
           Institute for Theoretical Physics, 
           Vienna University of Technology,\\
           Wiedner Hauptstr. 8--10/136,
           A-1040 Vienna, Austria. Europe\\
           Email: \email{afshar, mgary, grumil, rash, rieglerm@hep.itp.tuwien.ac.at}}

\abstract{
We present the general algorithm to establish the classical and quantum asymptotic symmetry algebra for non-AdS higher spin gravity and implement it for the specific example of spin-3 gravity in the non-principal embedding with Lobachevsky ($\mathbb{H}^2\times\mathbb{R}$) boundary conditions.
The asymptotic symmetry algebra for this example consists of a quantum $W_3^{(2)}$ (Polyakov--Bershadsky) and an affine $\hat u(1)$ algebra. 
We show that unitary representations of the quantum $W_3^{(2)}$ algebra exist only for two values of its central charge, the trivial $c=0$ ``theory'' and the simple $c=1$ theory. 
}


\preprint{TUW--12--22}

\begin{document}

\section{Introduction}

Higher spin \cite{Fradkin:1986qy} 
holography  \cite{Maldacena:1997re} has attracted considerable interest \cite{Mikhailov:2002bp} 
in the past few years \cite{Giombi:2009wh,Henneaux:2010xg,Campoleoni:2010zq,Gaberdiel:2010pz,Castro:2011fm,Creutzig:2011fe}.  
Recently, three of us suggested that 3-dimensional higher spin gravity \cite{Aragone:1983sz} 
can accommodate asymptotic backgrounds beyond AdS, and we provided evidence for this claim by explicitly constructing such backgrounds and showing compatibility with a well-defined variational principle \cite{Gary:2012ms}.
However, the litmus test of any such suggestion is the establishment of consistent boundary conditions that actually allow for these (non-AdS) backgrounds.


``Consistent'' here means that the canonical charges evaluated for these boundary conditions are finite, integrable and conserved. Once consistency is verified, the asymptotic symmetry algebra can be derived from the algebra of the canonical charges, which establishes the symmetry algebra of the dual field theory, including its central charges. This analysis therefore gives crucial clues about the dual field theory, like in the seminal work by Brown and Henneaux \cite{Brown:1986nw}, who showed that any consistent theory of 3-dimensional quantum gravity with asymptotically AdS boundary conditions is a 2-dimensional conformal field theory, in the sense that the Hilbert space falls into a representation of the conformal group in two dimensions. 

For AdS holography in 3-dimensional higher spin gravity the litmus test was provided in \cite{Henneaux:2010xg,Campoleoni:2010zq}.
In this paper we provide the litmus test for non-AdS holography in 3-dimensional higher spin gravity.

We study first general aspects and focus then on a specific example to spell out all the details, namely spin-3 gravity in the non-principal embedding with $\mathbb{H}^2\times\mathbb{R}$ boundary conditions, $\mathbb{H}^2$ being the Lobachevsky-plane (also known as ``Euklidean AdS$_2$''). We call these boundary conditions ``Lobachevsky boundary conditions''.

This paper is organized as follows.
In section \ref{se:2} we present the general algorithm for non-AdS higher spin holography. In the remainder of the paper we then implement the algorithm for the specific example.
In section \ref{se:3} we start with the background and propose boundary conditions.
In section \ref{se:4} we exploit the canonical analysis to check consistency of the boundary conditions and to derive the (semi-classical) asymptotic symmetry algebra.
In section \ref{se:5} we lift the semi-classical results to quantum results and finally end up with the Polyakov--Bershadsky $W_3^{(2)}$ algebra times an affine $\hat u(1)$ algebra.
In section \ref{se:6} we study unitary representations of this algebra and prove that they exist only for the trivial case $c=0$ and the simple case $c=1$.
In section \ref{se:7} we conclude and point out several generalizations. 

An extended introduction that explains the motivation to study non-AdS holography in 3-dimensional higher spin gravity and contains further references can be found in \cite{Gary:2012ms}. We use the same conventions as in \cite{Gary:2012ms} and set $\eps^{+-\rho}=-1$, where $x^\pm$ refers to the boundary coordinates and $\rho$ to the radial coordinate on the cylinder.

\section{General algorithm for non-AdS higher spin holography}\label{se:2}

The general algorithm essentially goes back to the work of Brown and Henneaux \cite{Brown:1986nw}.
\begin{enumerate}
 \item Identify bulk theory and variational principle
 \item Impose suitable boundary conditions 
 \item Perform canonical analysis and check consistency of boundary conditions
 \item Derive classical asymptotic symmetry algebra and its central charges
 \item Improve to quantum asymptotic symmetry algebra 
 \item Study unitary representations of the quantum asymptotic symmetry algebra
 \item Identify/constrain dual field theory
\end{enumerate}

\paragraph{1.} The first item for non-AdS holography in 3-dimensional higher spin gravity was resolved in \cite{Gary:2012ms}, whose main results we review now.
The bulk theory is always a Chern--Simons theory \cite{Achucarro:1987vz} 
whose gauge algebra contains $sl(2)\oplus sl(2)$, for instance $sl(n)\oplus sl(n)$. The theory is defined on some manifold $\mathcal{M}$ with boundary $\partial\mathcal{M}$ (usually we assume cylindrical topology $\mathcal{M}=\mathcal{D}\times\mathbb{R}$, so that the boundary is $S^1\times\mathbb{R}$).
\begin{equation}
		I=I_{\textrm{\tiny CS}}[A]-I_{\textrm{\tiny CS}}[\bar{A}]
\label{eq:nads1}
\end{equation}
with
\begin{equation}\label{Intro:ICS}
		I_{\textrm{\tiny CS}}[A]=\frac{k}{4\pi}\int_{\mathcal{M}}\textnormal{Tr}(A\wedge\extd A+
		\tfrac{2}{3}A\wedge A\wedge A)+B[A]
\end{equation}
and similarly for the connection $\bar A$, but with a relative minus sign in the boundary term $B[\bar A]$.
The Chern--Simons level $k$ is the only coupling constant of the theory.
The (gauge-invariant but diffeomorphism non-invariant) boundary term is given by\footnote{Equation \eqref{eq:angelinajolie} corrects a notational mistake in equations (2.9) and (2.10) of \cite{Gary:2012ms}.} 
\begin{equation}
		B[A]=\frac{k}{4\pi}\int_{\partial\mathcal{M}}\!\!\!\textnormal{Tr}(A_+\extd x^+\,A_-\extd x^-)
\label{eq:angelinajolie}
\end{equation}
where $x^\pm$ are some boundary coordinates (not necessarily light-cone coordinates).
The boundary term \eqref{eq:angelinajolie} is invariant under (anti-)holomorphic diffeomorphisms $x^\pm\to f^\pm(x^\pm)$. [In the AdS case where both boundary coordinates have the same scaling properties the boundary term \eqref{eq:angelinajolie} can be presented covariantly as $B[A]=\tfrac{k}{8\pi}\int_{\partial\mathcal{M}}\!\extd^2x\sqrt{-\ga}\,\ga^{ij}\,\textnormal{Tr}(A_i\,A_j)$, where $\ga$ is the induced metric on the boundary.]


Varying the action \eqref{Intro:ICS} with the boundary term \eqref{eq:angelinajolie} leads to flatness of the connection $A$ as the bulk equations of motion,
$F=0$,
where $F$ is the non-abelian field strength, and to boundary equations of motion that are solved by the boundary condition
\eq{
\delta A_-\big|_{\partial\mathcal{M}}=0 \qquad \textrm{or}\qquad A_+\big|_{\partial\mathcal{M}}=0\,.
}{eq:nads2}
The appropriate choice in \eqref{eq:nads2} depends on the specific example one is considering.
The bar-sector works analogously and can be obtained by exchanging $\pm$ components everywhere.
The bulk plus boundary action \eqref{eq:nads1}-\eqref{eq:angelinajolie} with the variational principle \eqref{eq:nads2} is suitable for generic non-AdS holography in 3-dimensional higher spin gravity.

\paragraph{2.} Part of the second item was also addressed in \cite{Gary:2012ms}, namely the existence and explicit construction of certain non-AdS backgrounds (Schr\"odinger, Lifshitz, warped AdS and Lobachevsky). It turned out that one has to pick a specific class of embeddings of $sl(2)$ into, say, $sl(n)$, depending on the desired background. For instance, the principal embedding can never produce general warped AdS backgrounds or Lobachevsky, because these backgrounds require the presence of a singlet, but they allow for Schr\"odinger and Lifshitz. The possible values of the Lifshitz/Schr\"odinger scaling exponent depend on the value of $n$. 
All non-principal embeddings have at least one singlet and thus allow a Lobachevsky background; most of these embeddings additionally allow for warped AdS backgrounds, with the possible values of the warping parameter depending on both $n$ and the specific embedding. The statements above were presented already in \cite{Gary:2012ms} to which we refer for further details. Let us just quote for later use the relation between the connections and the metric.
\eq{
g_{\mu\nu}=\frac12\,\textrm{Tr}\,\big[(A-\bar A)_\mu(A-\bar A)_\nu\big]
}{eq:nads7}
The combination $A-\bar A$ corresponds to the vielbein in the $sl(2)$ case and to the ``zuvielbein'' otherwise. Note that for non-principal embeddings the definition \eqref{eq:nads7} is not unique, since one can add singlet terms, see for instance \cite{Castro:2011fm}.
For the background metric it is usually convenient to employ Gaussian normal coordinates
\eq{
\extd s^2_{\textrm{\tiny BG}} = \extd\rho^2 + \ga_{ij}(x^+,\,x^-,\,\rho)\,\extd x^i \extd x^j
}{eq:nads8}
where $x^\pm$ are the ``boundary coordinates'' introduced above and $\rho$ is a ``radial'' coordinate. We assume that the asymptotic boundary is located at $\rho\to\infty$.

One part of the second item that was not addressed in \cite{Gary:2012ms} was the imposition of precise boundary conditions on the connections $A$ and $\bar A$, i.e., the fall-off behavior of the subleading contributions. We explain now how this works in general, and refer to section \ref{se:3} for a specific example. Take a suitable group element $b$ and write the connections as
\eq{
A_\mu = b^{-1} \,\big(\hat a_\mu^{(0)} + a_\mu^{(0)} + a_\mu^{(1)}\big)\, b \qquad  \bar A_\mu = b \,\big(\hat{\bar{a}}_\mu^{(0)} + \bar a_\mu^{(0)} + \bar a_\mu^{(1)}\big)\, b^{-1}\,.
}{eq:nads5}
``Suitable'' here means that the leading contributions $\hat a_\mu^{(0)}+ a_\mu^{(0)}$ and $\hat{\bar{a}}_\mu^{(0)} +\bar a_\mu^{(0)}$ can be chosen to be independent from the radial coordinate $\rho$ such that they generate the desired background discussed above (contained in the hatted part), together with some state-dependent contributions (contained in the unhatted part). 
A choice \cite{Banados:1994tn} 
that works in many cases is 
\eq{
b=e^{\rho\,L_0}
}{eq:nads14}
where $L_0$ is the $sl(2)$ Cartan generator. 

When choosing the background $\hat a_\mu^{(0)}$, $\hat{\bar{a}}_\mu^{(0)}$ [the leading state-dependent fluctuations $a_\mu^{(0)}$, $\bar a_\mu^{(0)}$] it is necessary [convenient] to impose the equations of motion $F=\bar F=0$ asymptotically. For the special case \eqref{eq:nads14} they are solved by \cite{Banados:1994tn,Coussaert:1995zp} $\hat a_\rho^{(0)}=L_0$, $a_\rho^{(0)}=0$ and
\eq{
\partial_+ \hat a_-^{(0)} = \partial_- \hat a_+^{(0)} \qquad \partial_+ a_-^{(0)} = \partial_- a_+^{(0)}\qquad \big[\hat a_+^{(0)}+ a_+^{(0)},\,\hat a_-^{(0)}+ a_-^{(0)}\big]=0
}{eq:nads16}
and similarly for the bar-sector. The conditions \eqref{eq:nads16} are sufficient but not necessary for asymptotic gauge-flatness. 
Asymptotic gauge-flatness imposes restrictions on the possible generators appearing in various components of the leading connection, and also restricts the coordinate-dependence of the state-dependent functions therein. This reduces some of the clutter in the Ansatz for the boundary conditions of the connections.

What remains to be done is to fix the subleading parts $a_\mu^{(1)}$ and $\bar a_\mu^{(1)}$. They are generically suppressed exponentially in $\rho$, but otherwise unrestricted; for instance, a typical fall-off condition is $a_\mu^{(1)}\sim \mathcal{O}(e^{-2\rho})$ and similarly for $\bar a_\mu^{(1)}$. 

The boundary conditions are preserved by gauge transformations with gauge parameter $\epsilon$ if the gauge variation produces only state-dependent or subleading components.
	\begin{equation}\label{AdSxR:BCPGT}
		\delta_\epsilon A^a_\mu=\partial_\mu\epsilon^a+f^a{}_{bc}A^b_\mu\epsilon^c=\mathcal{O}\big(b^{-1}a_\mu^{(0)} b\big)^a + \mathcal{O}\big(b^{-1} a_\mu^{(1)} b\big)^a
	\end{equation}
In order to find all such gauge transformations we expand the gauge parameter $\epsilon$ as
\eq{
\epsilon = b^{-1}\big(\epsilon^{(0)}+\epsilon^{(1)}\big)\,b
}{eq:nads17}
where $\epsilon^{(0)}$ is $\rho$-independent and $\epsilon^{(1)}$ is subleading. Choosing again \eqref{eq:nads14} and expanding the transformation law \eqref{AdSxR:BCPGT} into leading and subleading terms yields two conditions,
\eq{
\partial_\pm\epsilon^{(0)\,a} + f^a{}_{bc}\,\big(\hat a^{(0)}_\pm + a^{(0)}_\pm\big)^b \epsilon^{(0)\,c} = \mathcal{O}(a_\pm^{(0)})^a
}{eq:nads18}
and $\epsilon^{(1)} = {\cal O}\big(a^{(1)}\big)$.
The transformations generated by $\epsilon^{(0)}$ usually belong to the asymptotic symmetry group, while the transformations generated by $\epsilon^{(1)}$ are trivial gauge transformations that are modded out in the asymptotic symmetry group.

\paragraph{3.} The third item can be done in a background-independent way, i.e., in full generality. Following the standard procedure, see appendix \ref{app:A} and references therein, leads to the canonical gauge generators ${\cal G}$ in \eqref{eq:app2}, including a contribution corresponding to the the canonical boundary charges
\eq{
\delta Q[\epsilon]=\frac{k}{2\pi}\,\oint_{\partial\mathcal{D}}\!\!\Tr(\epsilon\,\delta A_\varphi\extd\varphi)\,.
}{eq:nads9}
The integral goes over the cycle of the boundary cylinder, which we coordinatize by $\varphi$.
Integrability of the charges \eqref{eq:nads9} in field space is manifest if the gauge parameters $\epsilon$ are field-independent. Otherwise integrability has to be checked. Finiteness of the charges \eqref{eq:nads9} follows easily from the asymptotic forms of connection \eqref{eq:nads5} and gauge parameter \eqref{eq:nads17}. Using cyclicity of the trace yields an expression that is manifestly $\rho$-independent in the large $\rho$ limit.
\eq{
\delta Q[\epsilon]=\frac{k}{2\pi}\,\oint_{\partial\mathcal{D}}\!\!\Tr(\epsilon^{(0)}\,\delta a^{(0)}_\varphi\extd\varphi)
}{eq:nads9a}
Conservation of the charges $Q[\epsilon]$ is not evident. This property must be checked on a case-by-case basis, which is very simple to do by inserting a given set of boundary conditions into the result \eqref{eq:nads9a} or its integrated version. In many cases conservation of the charges is evident since the gauge parameter $\epsilon^{(0)}$ and the leading order connection $a^{(0)}$ depend on the angular coordinate $\varphi$ only.

\paragraph{4.} Concerning the fourth item there is little we can say in full generality besides the obvious: One just has to work out the Dirac brackets between the gauge generators ${\cal G}$, and in this way one obtains the (semi-classical) asymptotic symmetry algebra, including the results for the central charges. There is a well-known short-cut that permits one to circumvent the direct calculation of Dirac brackets. Namely, suppose we have two gauge generators whose Dirac bracket $\{{\cal G}[\epsilon_1],\,{\cal G}[\epsilon_2]\}$ we want to evaluate. Then we exploit the fact that the canonical generators generate gauge transformations via the Dirac bracket, $\{{\cal G}[\epsilon_1],\,{\cal G}[\epsilon_2]\} = \de_{\epsilon_2}\,{\cal G}[\epsilon_1]$, and evaluate the variation of the gauge generator \eqref{eq:nads11} on the right hand side. In fact, this Dirac bracket relation is on-shell equivalent to a corresponding relation involving the boundary charges \eqref{eq:nads9}
\eq{
\{Q[\epsilon_1],\,Q[\epsilon_2]\} = \de_{\epsilon_2}\,Q[\epsilon_1]\,.
}{eq:nads11}
The right hand side of \eqref{eq:nads11} is usually easy to evaluate and directly leads to the (semi-classical) asymptotic symmetry algebra, including the (semi-classical) central terms.

\paragraph{5.} The fifth item did not arise in the Brown--Henneaux analysis \cite{Brown:1986nw}, but it appeared in the Henneaux--Rey \cite{Henneaux:2010xg} analysis. The key observation here is that the asymptotic symmetry algebra derived in the previous point may be valid only in the limit of large central charges. For moderate values of the central charges and taking into account normal ordering it can (and does) happen that the Jacobi identities are violated if one insists on the semi-classical results for the asymptotic symmetry algebra. The simplest way to address this is to allow suitable deformations of the semi-classical algebra, and to demand the validity of the Jacobi identities (see for instance \cite{Gaberdiel:2012ku}). The Jacobi identities then establish relations between the deformation parameters. Solving these relations eventually leads to the quantum asymptotic symmetry algebra. The only tricky part in this procedure may be the identification of what counts as ``suitable'' deformation. However, we believe that in most instances it should be clear which structure functions/structure constants must remain undeformed, and which can be allowed to be deformed. All this will be spelled out in detail for the specific example in section \ref{se:5}.

\paragraph{6.} Studying the sixth item follows general guidelines for establishing (non-)unitarity in (conformal) field theories. In particular, one has to check the non-negativity of the norm of all states in the spectrum. In general, this leads to a restriction of the possible values of the central charges appearing in the asymptotic symmetry algebra. This in turn implies a restriction on the Chern--Simons level $k$ appearing in the action \eqref{Intro:ICS}. Depending on the theory, it is possible to have infinitely many admissible values of $k$, some isolated values of $k$ or no value at all. In the last case, the holographic correspondence relates a non-unitary higher spin theory to a non-unitary field theory; while non-unitarity is often sufficient to drop the theory, there are some interesting applications involving non-unitary theories, for instance in the context of the AdS$_3$/log CFT$_2$ correspondence (see \cite{Grumiller:2008qz} 
and references therein).

\paragraph{7.} Finally, with the results from the first six items available, the seventh item consists mainly of putting all these clues together to restrict, or perhaps even uniquely identify, the dual field theory. Once a specific field theory is conjectured, further checks are possible, like the calculation of the partition function, checks of modular invariance, the calculation of correlators on the gravity side etc.~All these additional checks go beyond the scope of our present work.

\section{Background and boundary conditions}\label{se:3}

From now on we focus on a Lobachevsky background. 
\eq{
\extd s^2 = \extd t^2 +\extd\rho^2 + \sinh^2\!\rho\,\extd\varphi^2
}{eq:nads13}
For sake of specificity we choose Euklidean signature. In the conventions of section \ref{se:2} we identify the boundary coordinates as $t=x^+$, $\varphi=x^-$. The connections producing the background \eqref{eq:nads13} in the limit $\rho\to\infty$ are given by \cite{Gary:2012ms}.
\begin{subequations}\label{AdSxR:Connection}
	\begin{align}
		A_\rho={}&L_0&\bar{A}_\rho={}&-L_0\\
		A_\varphi={}&-\frac14\, e^\rho L_1&\bar{A}_\varphi={}&-e^\rho L_{-1}\\
		A_t={}&0&\bar{A}_t={}&\sqrt{3}S
	\end{align}
\end{subequations}
Besides the $sl(2)$ generators $L_n$ the background \eqref{eq:nads13} requires the presence of a singlet $S$ (with suitable normalizations). The simplest possibility is to consider the $sl(3)$ non-principal embedding, whose convention we summarize in appendix \ref{app:B}. The factors in \eqref{AdSxR:Connection} are adapted to these conventions. We have thus identified the bulk theory and the background solution, which fixes part of the first item in the algorithm at the beginning of section \ref{se:2}.

We fix now the background and fluctuation behavior by specifying the quantities $a^{(0,1)}$ and $\bar a^{(0,1)}$ in \eqref{eq:nads5} [and pick the group element $b$ as in \eqref{eq:nads14}].
	\begin{subequations}\label{AdSxR:BCs1}
		\begin{align}
			\hat a^{(0)}_\rho&=L_0 \quad \hat a^{(0)}_\varphi=-\frac{1}{4}\,L_1 \quad \hat{\bar{a}}^{(0)}_\rho=-L_0 \quad \hat{\bar{a}}^{(0)}_\varphi=-L_{-1} \quad \hat{\bar{a}}^{(0)}_t=\sqrt{3}S\\
			a^{(0)}_\varphi&=\frac{2\pi}{k}\,\big(\tfrac{3}{2}\mathcal{W}_0(\varphi)S+\mathcal{W}^{+}_{\frac{1}{2}}(\varphi)\psi^{+}_{-\frac{1}{2}}
				-\mathcal{W}^{-}_{\frac{1}{2}}(\varphi)\psi^{-}_{-\frac{1}{2}}-\mathcal{L}(\varphi)L_{-1}\big)\\
			\bar{a}^{(0)}_\varphi&=\frac{2\pi}{k}\,\big(\tfrac{3}{2}\bar{\mathcal{W}}_0(\varphi)S+\bar{\mathcal{W}}^+_{\frac12}(\varphi)\psi^{+}_{\frac{1}{2}}+ \bar{\mathcal{W}}^-_{\frac12}(\varphi)\psi^{-}_{\frac{1}{2}}+\bar{\mathcal{L}}(\varphi)L_1\big)\\
\hat a^{(0)}_t&=a^{(0)}_\rho=a^{(0)}_t=\bar a^{(0)}_\rho=\bar a_t^{(0)} = 0 \\
			a^{(1)}_\mu&=\mathcal{O}(e^{-2\rho})=\bar a^{(1)}_\mu
		\end{align}
	\end{subequations}
We explain now how we came up with the Ansatz \eqref{AdSxR:BCs1}. The leading components in $\hat a^{(0)}$ and $\hat{\bar{a}}^{(0)}$ are determined uniquely from the asymptotic background \eqref{AdSxR:Connection}. We employ here a mixed variational principle, with $A_t|_{\partial\cal M}=0$ and $\de\,\bar A_t|_{\partial\cal M}=0$. This explains why the state-dependent components $a^{(0)}_t$ and $\bar a^{(0)}_t$ vanish. We do not consider state-dependent components $a^{(0)}_\rho$ or $\bar a^{(0)}_\rho$, since they could violate asymptotic gauge-flatness. 
The fact that all arbitrary functions introduced in the Ansatz \eqref{AdSxR:BCs1} only depend on $\varphi$ ensures that the gauge-flatness conditions \eqref{eq:nads16} are solved. The state-dependent terms with arbitrary functions in $a^{(0)}_\varphi$ ($\bar a^{(0)}_\varphi$) contain all generators with non-positive (non-negative) weights, which leads to the desired fall-off behavior for the metric and, as we demonstrate in section \ref{se:5}, to integrability of the charges. The Ansatz for the subleading components in the last line of \eqref{AdSxR:BCs1} could be relaxed to weaker fall-off behavior as long as $\lim_{\rho\to\infty}a^{(1)}=0$.

As a consequence of our choices the ensuing metric \eqref{eq:nads7} obeys Lobachevsky boundary conditions \cite{warped}
\begin{equation}\label{AdSxR:MetricBoundary2}
		g_{\mu\nu}=\left(
			\begin{array}{ccc}
				g_{tt}=1+\mathcal{O}(e^{-2\rho})&g_{t\rho}=\mathcal{O}(e^{-2\rho})& g_{t\varphi}=\mathcal{O}(1)\\
				&g_{\rho\rho}=1+\mathcal{O}(e^{-2\rho})&g_{\rho\varphi}=\mathcal{O}(1)\\
				&&g_{\varphi\varphi}=\tfrac14\, e^{2\rho}+\mathcal{O}(1)
			\end{array}\right).
\end{equation}
At this stage the first and second item in section \ref{se:2} are dealt with, except for showing that the boundary conditions \eqref{AdSxR:BCs1} are suitable. While a full proof of suitability must await the canonical analysis in section \ref{se:4}, as a first step in this direction we consider the gauge transformations that preserve the boundary conditions \eqref{AdSxR:BCs1} and show that they are non-trivial.

To this end we decompose the leading contribution in \eqref{eq:nads17} into components, ordered by their $sl(2)$ weights,
	\begin{equation}
		\epsilon^{(0)}=\epsilon_1\,L_1+\epsilon_{\frac12}^+\,\psi^{+}_{\frac{1}{2}}+\epsilon_{\frac12}^-\,\psi^{-}_{\frac{1}{2}}+\epsilon_0^L\,L_0+\epsilon_0^S\,S+\epsilon_{-\frac12}^+\,\psi^{+}_{-\frac{1}{2}}+
		\epsilon_{-\frac12}^-\,\psi^{-}_{-\frac{1}{2}}+\epsilon_{-1}\,L_{-1}\,.
\label{eq:nads19}
	\end{equation}
Solving the condition \eqref{eq:nads18} with the decomposition \eqref{eq:nads19} yields (prime denotes $\varphi$-derivatives)
	\begin{subequations}\label{AdSxR:BCPGTASectorDetail}
		\begin{align}
			\epsilon_1&=\epsilon(\varphi)\qquad\epsilon_{\frac12}^\pm=\epsilon^\pm_{\frac{1}{2}}(\varphi)\qquad\epsilon_0^L=4\epsilon'(\varphi)\qquad\epsilon_0^S=\epsilon_0(\varphi)\\
			\epsilon_{-\frac12}^\pm&=4{\epsilon^\pm_{\frac{1}{2}}}'(\varphi)\mp\frac{4\pi}{k}
				\Big(2\mathcal{W}^\pm_{\frac{1}{2}}(\varphi)\epsilon(\varphi)-3\mathcal{W}_0(\varphi)\epsilon^\pm_{\frac{1}{2}}(\varphi)\Big)\\
			\epsilon_{-1}&=8\epsilon''(\varphi)+\frac{4\pi}{k}\Big(2\mathcal{L}(\varphi)\epsilon(\varphi)+
				\mathcal{W}^{-}_{\frac{1}{2}}(\varphi)\epsilon^{+}_{\frac{1}{2}}(\varphi)+
				\mathcal{W}^{+}_{\frac{1}{2}}(\varphi)\epsilon^{-}_{\frac{1}{2}}(\varphi)\Big)\,.
		\end{align}
	\end{subequations}
It is worthwhile noting the pattern that emerges here: the coefficient functions $\epsilon_q$ with positive $q$ are all free functions of $\varphi$, while the ones with negative $q$ are all uniquely determined by these functions, but in a way that is state-dependent. 

The non-trivial gauge transformation that preserve the boundary conditions 
are then parametrized by four free functions of the coordinate $\varphi$. It is convenient to introduce the notation
\eq{
\delta_{\epsilon^{(0)}} = \delta_{\epsilon}+\delta_{\epsilon_0}+\delta_{\epsilon^{+}_{\frac{1}{2}}}+\delta_{\epsilon^{-}_{\frac{1}{2}}}\,.
}{eq:nads20}
The transformation $\delta_{\epsilon}$ is a gauge transformation 
where all free functions are set to zero, except for the function $\epsilon(\varphi)$ (and similarly for $\delta_{\epsilon_0}$ and $\delta_{\epsilon^\pm_{\frac{1}{2}}}$). 
Let us finally note that $\delta_\epsilon$ generates (holomorphic) diffeomorphisms. Indeed, it is possible to write the gauge parameter essentially as $\epsilon^a=\xi^\mu\,A_\mu^a$ with $\xi^t=0$, $\xi^\varphi=-4\epsilon(\varphi)$ and $\xi^\rho=4\epsilon'(\varphi)$.\footnote{\label{fn:1} The only two obstructions to writing $\epsilon^a$ in that form are the standard $\epsilon''(\varphi)$-term in the component $\epsilon_{-1}$ that ultimately is responsible for the Virasoro central charge, and a term of the form $-\tfrac{12\pi}{k}\,\epsilon(\varphi)\mathcal{W}_0(\varphi)$ in the component $\epsilon_0^S$. The latter term can be absorbed by a state-dependent redefinition of the function $\epsilon_0(\varphi)$. This is taken care of precisely by the Sugawara-shift \eqref{eq:sugawara} below and leads to an integrable shift of the canonical charges by $\delta Q^{\rm shift}=-\tfrac{12\pi}{k}\,\oint\extd\varphi\,\epsilon\mathcal{W}_0\delta\mathcal{W}_0$.}

\section{Canonical analysis and asymptotic symmetry algebra}\label{se:4}

The canonical analysis was performed in full generality in section \ref{se:2} and appendix \ref{app:A}. To check consistency of the boundary conditions we just have to plug the boundary-condition preserving gauge transformations \eqref{eq:nads19}, \eqref{AdSxR:BCPGTASectorDetail} and the variation of the connection \eqref{eq:nads5} with our boundary conditions \eqref{AdSxR:BCs1} into the general result for the canonical charges \eqref{eq:nads9} and verify whether the charges are integrable in field space and conserved in time [finiteness was already checked in section \ref{se:2} in full generality, see \eqref{eq:nads9a}].

In the present case the variation of the boundary charges \eqref{eq:nads9a} can be integrated in field-space, since the whole state-dependence of the gauge parameter $\epsilon^{(0)}$ \eqref{eq:nads19}, \eqref{AdSxR:BCPGTASectorDetail} resides in the negative weight generators, which would have to combine with positive weight state dependence in the variation of the connection $\delta a^{(0)}$ to give a non-vanishing trace. As the positive weight part of the connection \eqref{AdSxR:BCs1} does not contain any state dependence, the integrated charges are simply given by
	\begin{equation}\label{AdSxR:BoundaryChargeFinal}
		Q[\epsilon^{(0)}]=\oint\extd\varphi\,\big(\mathcal{L}\epsilon+\mathcal{W}_0\epsilon_0+
		\mathcal{W}_{\frac{1}{2}}^{+}\epsilon_{\frac{1}{2}}^{-}+\mathcal{W}_{\frac{1}{2}}^{-}\epsilon_{\frac{1}{2}}^{+}\big)\,.
	\end{equation}
So far we have considered only the $A$-sector; the $\bar A$-sector works in full analogy, and yields the charges
\begin{equation}
	\bar{Q}[\bar{\epsilon}^{(0)}]=-\oint\extd\varphi\,\bar{\mathcal{W}}_0\bar{\epsilon}_0
\label{eq:barQ}
\end{equation}
with 
\begin{equation}
	\bar{\epsilon}^{(0)}=\bar{\epsilon}_0(\varphi)\,S\,.
\end{equation}
Evidently, the $\bar A$-sector is much simpler than the $A$-sector.
We have shown in section \ref{se:2} that the charges are finite, and above that they are integrable. From the expressions \eqref{AdSxR:BoundaryChargeFinal} and \eqref{eq:barQ} it is evident that they are conserved. Thus, we can tick off items two and three from the list in section \ref{se:2}.

The asymptotic symmetry algebra follows from the Dirac bracket algebra of the canonical generators. The short-cut \eqref{eq:nads11} requires the evaluation of the variations of the canonical charges \eqref{AdSxR:BoundaryChargeFinal}. Therefore, we consider now the variation of the state-dependent functions $\mathcal{L}$, $\mathcal{W}^\pm_{\frac{1}{2}}$ and $\mathcal{W}_0$ under the boundary-condition preserving gauge transformations \eqref{eq:nads20}. We obtain
	\begin{subequations}\label{AdSxR:Trafos}
		\begin{align}
			&\delta_{\epsilon}\mathcal{L}=-4\big(2\epsilon'\mathcal{L}+
			\epsilon\mathcal{L}'\big)-\frac{4k}{\pi}\epsilon'''\qquad\delta_{\epsilon}\mathcal{W}_0=0\\
			&\delta_{\epsilon}\mathcal{W}^{\pm}_{\frac{1}{2}}=-4\big(\frac{3}{2}\epsilon'
			\mathcal{W}^{\pm}_{\frac{1}{2}}+\epsilon{\mathcal{W}^{\pm}_{\frac{1}{2}}}'+
			\pm\frac{3\pi}{k}\epsilon\mathcal{W}^{\pm}_{\frac{1}{2}}\mathcal{W}_0\big)\label{AdSxR:TrafosEps}\\
			&\delta_{\epsilon_0}\mathcal{L}=0\qquad\delta_{\epsilon_0}\mathcal{W}_0=\frac{k}{3\pi}\epsilon_0'\qquad
			\delta_{\epsilon_0}\mathcal{W}^{\pm}_{\frac{1}{2}}=
			\mp\epsilon_0\mathcal{W}^{\pm}_{\frac{1}{2}}\\
&\delta_{\epsilon^{\pm}_{\frac{1}{2}}}\mathcal{L}=-2\big(
			\pm\frac{6\pi}{k}\mathcal{W}_0\mathcal{W}_{\frac{1}{2}}^{\mp}\epsilon_{\frac{1}{2}}^{\pm}+
			{\mathcal{W}_{\frac{1}{2}}^{\mp}}'\epsilon_{\frac{1}{2}}^{\mp}+3\mathcal{W}_{\frac{1}{2}}^{\mp}{\epsilon_{\frac{1}{2}}^{\pm}}'\big)\qquad
			\delta_{\epsilon^{\pm}_{\frac{1}{2}}}\mathcal{W}_0=\mp
			\mathcal{W}^{\mp}_{\frac{1}{2}}
			\epsilon_{\frac{1}{2}}^{\pm}\\
			&\delta_{\epsilon^{\pm}_{\frac{1}{2}}}\mathcal{W}^{\pm}_{\frac{1}{2}}=\pm\epsilon^{\pm}_{\frac{1}{2}}\mathcal{L}+4\big(3{\epsilon^{\pm}_{\frac{1}{2}}}'\mathcal{W}_0+\frac{3}{2}\epsilon^{\pm}_{\frac{1}{2}}{\mathcal{W}_0}'\pm
			\frac{9\pi}{2k}\epsilon^{\pm}_{\frac{1}{2}}\mathcal{W}_0\mathcal{W}_0\pm\frac{k}{2\pi}{\epsilon^{\pm}_{\frac{1}{2}}}''\big)
		\end{align}
	\end{subequations}
and $\delta_{\epsilon^{\pm}_{\frac{1}{2}}}\mathcal{W}^{\mp}_{\frac{1}{2}}=0$.
The first line in \eqref{AdSxR:Trafos} shows that the state-dependent function $\mathcal{L}$ transforms like the stress-energy tensor in a CFT, including the anomalous term proportional to the central charge. This is a good indication that we are on the right track. 

The transformations \eqref{AdSxR:Trafos} contain already the relevant information about the asymptotic symmetries. We split the canonical charges into individual contributions: $Q[\epsilon^{(0)}=\epsilon]=\oint\extd\varphi\,\epsilon {\cal L}$, $Q[\epsilon^{(0)}=\epsilon_{\frac12}^\pm]=\oint\extd\varphi\,\epsilon_{\frac12}^\pm{\cal W}^\pm_{\frac12}$, $Q[\epsilon^{(0)}=\epsilon_0]=\oint\extd\varphi\,\epsilon_0{\cal W}_0$. Exploiting the short-cut \eqref{eq:nads11} and the transformation formulas \eqref{AdSxR:Trafos} we then obtain the asymptotic symmetry algebra in a suggestive form.
	\begin{subequations}\label{AdSxR:W32AlgebraDirac}
		\begin{align}
&\{\mathcal{L}(\varphi),\mathcal{L}(\bar{\varphi})\}=-4\big(2\mathcal{L}\delta'(\varphi-\bar{\varphi})
			-\mathcal{L}'\delta(\varphi-\bar{\varphi})\big)-\frac{4k}{\pi}\delta'''(\varphi-\bar{\varphi})\label{AdSxR:W32AlgebraDiracLL}\\
			&\{\mathcal{L}(\varphi),\mathcal{W}_0(\bar{\varphi})\}=0\label{AdSxR:W32AlgebraDiracb}\\
&\{\mathcal{L}(\varphi),\mathcal{W}_{\frac{1}{2}}^{\pm}(\bar{\varphi})\}=
			-4\Big(\frac{3}{2}\mathcal{W}_{\frac{1}{2}}^{\pm}\delta'(\varphi-\bar{\varphi})-
			\big({\mathcal{W}_{\frac{1}{2}}^{\pm}}'\pm\frac{3\pi}{k}\mathcal{W}_{\frac{1}{2}}^{\pm}
			\mathcal{W}_0\big)\delta(\varphi-\bar{\varphi})\Big)\label{AdSxR:W32AlgebraDiracd}\\
			&\{\mathcal{W}_0(\varphi),\mathcal{W}_0(\bar{\varphi})\}=\frac{k}{3\pi}\delta'(\varphi-\bar{\varphi})\\
			&\{\mathcal{W}_0(\varphi),\mathcal{W}_{\frac{1}{2}}^{\pm}(\bar{\varphi})\}=
			\pm\mathcal{W}_{\frac{1}{2}}^{\pm}\delta(\varphi-\bar{\varphi})\\
			&\{\mathcal{W}_{\frac{1}{2}}^{+}(\varphi),\mathcal{W}_{\frac{1}{2}}^{-}(\bar{\varphi})\}=
			\mathcal{L}\delta(\varphi-\bar{\varphi})-4\Big(-3\mathcal{W}_0\delta'(\varphi-\bar{\varphi})+
			\big(\frac{3}{2}{\mathcal{W}_0}'-\frac{9\pi}{2k}\mathcal{W}_0\mathcal{W}_0\big)\nonumber\\
			&\qquad\qquad\qquad\qquad \qquad\qquad\qquad\qquad
			\delta(\varphi-\bar{\varphi})-
			\frac{k}{2\pi}\delta''(\varphi-\bar{\varphi})\Big)
		\end{align}
	\end{subequations}
This algebra is written in a non-primary basis as one can see by looking at \eqref{AdSxR:W32AlgebraDiracb} and \eqref{AdSxR:W32AlgebraDiracd}. This can be fixed by a Sugawara-shift of $\mathcal{L}$
	\begin{equation}
		\mathcal{L}\rightarrow\mathcal{L}-\frac{6\pi}{k}\mathcal{W}_0\mathcal{W}_0\equiv\hat{\mathcal{L}}\,.
\label{eq:sugawara}
	\end{equation}
The Sugawara-shift \eqref{eq:sugawara} emerges automatically from the redefinition of the function $\epsilon_0(\varphi)$ mentioned in footnote \ref{fn:1} at the end of section \ref{se:3}.

Replacing Dirac brackets by commutators, $i\{\cdot,\cdot\}\rightarrow[\cdot,\cdot]$, taking into account the Sugawara-shift \eqref{eq:sugawara}, and introducing Fourier modes for all generators
\begin{subequations}
\begin{align}
 \mathcal{L}(\varphi)&=-\frac{4}{2\pi}\sum_{n\in\mathbb{Z}}L_ne^{-in\varphi}
		&\hat{L}_n&= L_n-\frac{3}{4k}\sum_{p\in\mathbb{Z}}J_{n-p}J_p\\
\mathcal{W}_0(\varphi)&=\frac{i}{2\pi}\sum_{n\in\mathbb{Z}}J_ne^{-in\varphi}\qquad
		&\mathcal{W}_{\frac{1}{2}}^{\pm}(\varphi)&=\frac{(-4i)^{\frac{1\mp1}{2}}}{2\pi}\sum_{n\in\mathbb{Z}+\frac{1}{2}}G_n^{\pm}e^{-in\varphi}
\end{align}
\end{subequations}
leads to the desired commutator algebra, the $A$-sector of the asymptotic symmetry algebra.
	\begin{subequations}\label{AdSxR:W32AlgebraShifted}
		\begin{align}
			&[J_n,J_m]=-\frac{2k}{3}n\delta_{n+m,0}\\
			&[J_n,\hat{L}_m]=nJ_{n+m}\\
			&[J_n,G_m^{\pm}]=\pm G_{m+n}^{\pm}\\
			&[\hat{L}_n,\hat{L}_m]=(n-m)\hat{L}_{m+n}+\frac{c}{12}n\big(n^2-1\big)\delta_{n+m,0}\\
			&[\hat{L}_n,G_m^{\pm}]=\big(\frac{n}{2}-m\big)G_{n+m}^{\pm}\\
			&[G_n^{+},G_m^{-}]=\hat{L}_{m+n}-\frac{3}{2}(n-m)J_{m+n}+\frac{3}{k}\sum_{p\in\mathbb{Z}}J_{m+n-p}J_p+k\big(n^2-\frac{1}{4}\big)\delta_{m+n,0}\label{AdSxR:W32AlgebraShifted6}
		\end{align}
	\end{subequations}
The algebra \eqref{AdSxR:W32AlgebraShifted} is the semi-classical (large $k$) limit of the Polyakov--Bershadsky algebra, denoted by $W_3^{(2)}$.
The semi-classical Virasoro central charge is given by $c=6k$.
For later purposes we note that the normalization of the Virasoro generators $L_n$ and $R$-current generators $J_n$ is fixed by the commutators above, but the normalization of the symmetry-generators $G^\pm_n$ is arbitrary if we are willing to rescale the structure constants in the last commutator \eqref{AdSxR:W32AlgebraShifted6}.

The $\bar A$-sector is much simpler. 
The Dirac brackets for the $\bar{A}$ sector are given by
	\begin{equation}
		\{\bar{\mathcal{W}}_0(\varphi),\bar{\mathcal{W}}_0(\bar{\varphi})\}=-\frac{k}{3\pi}\delta'(\varphi-\bar{\varphi})\,.
\label{eq:nads23}
	\end{equation}
Plugging the Fourier modes
\eq{
\bar{\mathcal{W}_0}(\varphi)=\frac{1}{2\pi}\sum_{n\in\mathbb{Z}}\bar{J}_ne^{-in\varphi}
}{eq:nads22}
into \eqref{eq:nads23} yields an affine $\hat u(1)$ algebra.
\eq{
[\bar J_n,\bar J_m]=-\frac{2k}{3}\,n\delta_{n+m,0}
}{eq:nads21}

In conclusion, we have found above that the (semi-classical) asymptotic symmetry algebra consists of one copy of the (semi-classical) $W_3^{(2)}$ algebra \eqref{AdSxR:W32AlgebraShifted} and an affine $\hat u(1)$ algebra \eqref{eq:nads21}. This completes the fourth item of section \ref{se:2}.

\section{Quantum asymptotic symmetry algebra}\label{se:5}

Since we are interested in the quantum mechanical version of the asymptotic symmetry algebra \eqref{AdSxR:W32AlgebraShifted} we take into account normal ordering.
	\begin{equation}
		\sum_{p\in\mathbb{Z}}:J_{n-p}J_p:=\sum_{p\geq0}J_{n-p}J_p+\sum_{p<0}J_pJ_{n-p}
	\end{equation}
The normal ordered version of the algebra \eqref{AdSxR:W32AlgebraShifted} then becomes inconsistent for finite values of $k$ and thus has to be modified.
Therefore, we introduce five arbitrary coefficients $C_{1..5}$ and start with the Ansatz
	\begin{subequations}\label{AdSxR:W32AlgebraQuantum}
		\begin{align}
			&[J_n,J_m]=-C_1\,\frac{2k}{3}\,n\delta_{n+m,0}\\
			&[G_n^{+},G_m^{-}]=C_2\,\hat{L}_{m+n}-C_3\,\frac{3}{2}(n-m)J_{m+n}\nonumber\\
&\qquad\qquad\quad+C_4\,\frac3k\sum_{p\in\mathbb{Z}}:J_{m+n-p}J_p:+C_5\, k(n^2-\frac{1}{4})\delta_{m+n,0}\label{AdSxR:W32AlgebraQuantum6}\,.
		\end{align}
	\end{subequations}
The commutators not given in \eqref{AdSxR:W32AlgebraQuantum} are identical to the semi-classical ones in \eqref{AdSxR:W32AlgebraShifted}, which we do not want to deform. For $C_i=1$ the semi-classical algebra \eqref{AdSxR:W32AlgebraShifted} is recovered. Since we allow a deformation of \eqref{AdSxR:W32AlgebraQuantum6} we can additionally permit rescalings of the symmetry generators $G_n^\pm$, a freedom that we shall exploit below to map our results into standard form.

The coefficients $C_i$ are constrained by the Jacobi identities involving at least one symmetry generator $G^\pm$, which leads to four linear relations in the five parameters $C_i$ that also involve the central charge $\hat c$. So we have four equations in six variables and can fix two of them. To match with semi-classical results a viable choice of coefficients is $C_1=C_2=1$.
The Jacobi identities then yield the following coefficients and shifted central charge $C_3=1+\frac{8}{4k-6}$, $C_4=\frac{4k}{4k-6}$, $C_5=1+\frac{8}{4k-6}$, $\hat{c}=\frac{32k}{2k-3}+6k$. This is essentially the quantum $\mathcal{W}_3^{(2)}$ algebra found by Polyakov and Bershadsky in \cite{Polyakov:1989dm} but with a different $k$ and different normalization of the spin-$\frac{3}{2}$ modes. In order to bring this algebra in a more familiar form we apply the following shift of $k$ and renormalization of $G_n^{\pm}$
	\begin{equation}\label{AdSxR:KGShift}
		\hat k=-\big(k+\tfrac32\big)\qquad \hat{G}_n^{\pm}=G_n^{\pm}\sqrt{k-\tfrac32}\,.
	\end{equation}
This results in the standard form of the Polyakov--Bershadsky algebra
	\begin{subequations}\label{AdSxR:W32AlgebraQuantumBershadsky}
		\begin{align}
			&[J_n,J_m]=\frac{2\hat{k}+3}{3}n\delta_{n+m,0}\label{AdSxR:W32AlgebraQuantumBershadsky1}\\
			&[J_n,\hat{L}_m]=nJ_{n+m}\\
			&[J_n,\hat{G}_m^{\pm}]=\pm \hat G_{m+n}^{\pm}\\
			&[\hat{L}_n,\hat{L}_m]=(n-m)\hat{L}_{m+n}+\frac{\hat{c}}{12}n(n^2-1)\delta_{n+m,0}\\
			&[\hat{L}_n,\hat{G}_m^{\pm}]=\left(\frac{n}{2}-m\right)\hat{G}_{n+m}^{\pm}\\
			&[\hat{G}_n^{+},\hat{G}_m^{-}]=-(\hat{k}+3)\hat{L}_{m+n}+\frac{3}{2}(\hat{k}+1)(n-m)J_{m+n}
			+3\sum_{p\in\mathbb{Z}}:J_{m+n-p}J_p:\nonumber\\
			&\qquad\qquad\quad +\frac{(\hat{k}+1)(2\hat{k}+3)}{2}(n^2-\frac{1}{4})\delta_{m+n,0}\label{AdSxR:W32AlgebraQuantumBershadsky6}
		\end{align}
	\end{subequations}
with the standard result for the central charge
	\begin{equation}
		\hat{c}=25-\frac{24}{\hat{k}+3}-6(\hat{k}+3)=-\frac{(2\hat{k}+3)(3\hat{k}+1)}{\hat{k}+3}\,.
\label{eq:c}
	\end{equation}
The maximum value of the central charge, $\hat c=1$, is obtained for $\hat k=-1$. Non-negativity of $\hat c$ requires the level $\hat k$ to lie in the interval $-\tfrac13\geq\hat k\geq-\tfrac32$. These inequalities exclude the possibility of a unitary field theory dual in the semi-classical limit $|\hat k|\to\infty$ \cite{Castro:2012bc}.

In conclusion, the quantum asymptotic symmetry algebra consists of one copy of the (quantum) $W_3^{(2)}$ algebra \eqref{AdSxR:W32AlgebraQuantumBershadsky} with central charge \eqref{eq:c} and an affine $\hat u(1)$ algebra \eqref{eq:nads21}. This completes the fifth item of section \ref{se:2} and may have been anticipated on general grounds: after all, AdS$_3$ holography in the non-principal embedding of spin-3 gravity leads to two copies of the (quantum) $W_3^{(2)}$ algebra \cite{Ammon:2011nk}, and Lobachevsky holography respects half of the AdS isometries plus an extra abelian isometry, so the breaking of $W_3^{(2)}\oplus W_3^{(2)}\to W_3^{(2)}\oplus\hat u(1)$ appears quite natural.

\section{Unitary representations and possible field theory duals}\label{se:6}


The $W_3^{(2)}$ algebra \eqref{AdSxR:W32AlgebraQuantumBershadsky} resembles a bosonic version of the $N=2$ super-conformal algebra, with Virasoro generators $\hat L_n$, pseudo-supersymmetry generators $\hat G^\pm_n$, and ``$R$-current'' generators $J_n$. The generators, being purely bosonic, all satisfy
commutation relations. In particular, the $\hat G^\pm$ satisfy commutation relations, as opposed to anti-commutation relations as in
the $N=2$ super-conformal algebra; this strongly impacts the set of possible unitary representations of the algebra. The
purpose of this section is to study these representations, thus fulfilling the sixth item on the list in section \ref{se:2}.

Since the $\bar{A}$-sector only consists of an affine $\hat u(1)$ algebra \eqref{eq:nads21} it is not hard to find unitary representations. We define our vacuum by the conditions
\eq{
\bar J_n |0\rangle = 0
}{eq:nads25}
for all non-negative integer $n$.
Non-negativity of the physical states requires non-positive $k$ or, equivalently,
\eq{
\hat k \geq -\frac32\,.
}{eq:nads24}

The $A$-sector is more interesting. We define the vacuum state by 
	\begin{equation}\label{AdSxR:AnnihilationWeight}
		\hat L_{n-1}|0\rangle=0\qquad J_n|0\rangle=0\qquad \hat{G}^\pm_{n-\frac12}|0\rangle=0
	\end{equation} 
for all non-negative integer $n$.
The hermitian conjugates are defined by
	\begin{equation}\label{AdSxR:AllConjugate}
		\big(\hat L_n\big)^\dagger\equiv \hat L_{-n}\qquad\big(J_n\big)^\dagger\equiv J_{-n}\qquad\big(\hat{G}^\pm_n\big)^\dagger\equiv \hat{G}^\mp_{-n}\,.
	\end{equation}
We check now the norm of descendants of the vacuum in order to derive restrictions on the level $k$, similar (at least in spirit) to the inequality \eqref{eq:nads24}.
At level 1 there is only the state $J_{-1}|0\rangle$, whose norm 
	\begin{equation}\label{AdSxR:Level1Norm}
		\langle0|J_1J_{-1}|0\rangle=\frac{2\hat{k}+3}{3}
	\end{equation}
is already non-negative if the inequality \eqref{eq:nads24} holds.
The level-$\frac{3}{2}$ states $\hat{G}^\pm_{-\frac{3}{2}}|0\rangle$ lead to a Gramian matrix 
	\begin{equation}\label{AdSxR:LevelOnehalfNorm}
		K^{(\frac{3}{2})}=(\hat{k}+1)(2\hat{k}+3)\left(
		\begin{array}{cc}
			-1&0\\
			0&1					
		\end{array}\right)
	\end{equation}
that has a positive and a negative eigenvalue, unless the overall coefficient vanishes.
This means, already at this level there is always a positive and a negative norm state, except when the pre-factor in \eqref{AdSxR:LevelOnehalfNorm} vanishes.
This (significant) feature is in stark contrast to the $N=2$ super-conformal algebra. The relative minus sign in the matrix entries \eqref{AdSxR:LevelOnehalfNorm} is a direct consequence of the fact that the last entry in the algebra \eqref{AdSxR:W32AlgebraQuantumBershadsky6} is a commutator, rather than an anti-commutator. 

We conclude that for our choice of vacuum there are only two values of the level $\hat k$ where unitary representations of the $W_3^{(2)}$ algebra are possible, $\hat k=-\frac32$ or $\hat k=-1$. We discuss each of these possibilities in turn, in order to address the last remaining item seven on the list in section \ref{se:2}.

\paragraph{$\hat{k}=-\frac{3}{2}$ and $\hat{c}=0$}
This case is consistent but trivial: The only state present in our Hilbert space is the vacuum state itself. 

\paragraph{$\hat{k}=-1$ and $\hat{c}=1$}
For this value all half integer valued levels contain only null states \cite{Riegler:2012}, $\hat G^\pm_n|0\rangle=0$ for all possible $n$. Thus also the right hand side of the commutator \eqref{AdSxR:W32AlgebraQuantumBershadsky6} has to be zero. This leads to a relation
	\begin{align}\label{AdSxR:LasLinearCombofJ}
		\hat{L}_{-n}|0\rangle=\frac{3}{2}\,\sum_{p>0}^{n-1}J_{-p}J_{-n+p}|0\rangle=\frac{3}{2}\sum_{p\in\mathbb{Z}}:J_{-p}J_{-n+p}:|0\rangle
	\end{align}
that allows to express the Virasoro descendants of the vacuum, $\hat{L}_{-n}|0\rangle$, as a combination of $J$-descendants of the vacuum, which simplifies the theory considerably. It is now easy to check that all the remaining states in the theory, $J_{-n_1}^{m_1}\ldots J_{-n_N}^{m_N}|0\rangle$, have positive norm.
The resulting unitary CFT looks very similar to a (non-supersymmetric) theory based on an affine $\hat u(1)$ algebra 
	\begin{equation}\label{AdSxR:JJCommutatorU(1)}
		[{\cal J}_n,{\cal J}_m]=n\kappa\delta_{m+n,0}
	\end{equation}
and the Sugawara construction for the Virasoro modes ${\cal L}_n=\frac{1}{2\kappa}\sum_{p\in\mathbb{Z}}:{\cal J}_{-n-p}{\cal J}_p:$.
The crucial difference to our case is that in the algebra \eqref{AdSxR:JJCommutatorU(1)} the level $\kappa$ and the normalization of the generators ${\cal J}$ are both arbitrary. In our case, the normalization of the $R$-current is still fixed by the presence of the symmetry generators $\hat G^\pm$, and the level $\hat k$  is restricted to a unique value $\hat k=-1$, which corresponds to choosing $\kappa=\frac13$ in \eqref{AdSxR:JJCommutatorU(1)}.
Thus, even though the generators $\hat G^\pm$ do not lead to non-trivial physical states, their presence in the quantum asymptotic symmetry algebra is still noted by the level $\hat k$. 

\paragraph{Summary}
The only non-trivial unitary field theory dual to spin-3 gravity in the non-principal embedding with Lobachevsky boundary conditions emerges for a unique choice of the level, $\hat k=-1$ (or, equivalently, $k=-\frac12$).
The asymptotic symmetry algebra contains two affine $\hat u(1)$ algebras with the same value of the level, one coming from the $\bar A$-sector and one coming from the $A$-sector. The dual CFT has central charge $\hat c=1$ and corresponds to a free boson with a specific normalization of the boson's kinetic term.

\section{Conclusions}\label{se:7}

In this paper we confirmed the suggestion \cite{Gary:2012ms} that non-AdS holography in 3-di\-men\-sio\-nal higher spin gravity is possible, first by studying generic features of such holographic correspondences and then by providing an explicit example in the non-principal embedding of spin-3 gravity with Lobachevsky boundary conditions. The asymptotic symmetry algebra for this example contains an affine $\hat u(1)$ and the Pol\-ya\-kov--Ber\-shads\-ky algebra $W_3^{(2)}$. We confirmed the result \cite{Castro:2012bc} that there are no unitary representations of this algebra in the semi-classical limit, thus continuing a trend that seems fairly generic in 3-dimensional pure gravity theories \cite{Castro:2011zq}. We also found two values of the level $\hat k$ for which the algebra has unitary representations, one of which was trivial (vanishing central charge, $\hat c=0$). The other option (unity central charge, $\hat c=1$) led to the identification of the dual CFT as a theory of a single free boson. Therefore, we conclude that spin-3 gravity with Lobachevsky boundary conditions at $\hat c=1$ is dual to a free boson.

It is possible that there is another CFT groundstate with vacuum conditions different from \eqref{eq:nads25}, \eqref{AdSxR:AnnihilationWeight}. Such a groundstate may allow for more/other values of the level $\hat k$ leading to unitary representations of the Polyakov--Bershadsky algebra \eqref{AdSxR:W32AlgebraQuantumBershadsky}. It is not clear what happens on the gravity side in this case.

Several generalizations are possible, some of which are straightforward, but nonetheless potentially very interesting.
Imposing Lobachevsky boundary conditions in spin-$n$ gravity might lead to either of these possibilities: 1.~There are fewer/no values of the level $k$ where the theory becomes unitary; 2.~There are more values of the level $k$ where the theory becomes unitary, and the number of these values grows in some way with $n$; 3.~There are always ${\cal O}(1)$ values of $k$ where the theory becomes unitary. The next-simplest case is the 2-1-1 embedding in spin-4 gravity. More generally, one could study $sl(3)$ embeddings into $sl(n)$, where the $sl(3)$ contains the non-principally embedded $sl(2)$ factor. 
%
Such a comprehensive study could perhaps answer the question whether non-unitarity at large (but fixed) level $k$ is generic. In other words, given a (suitable rational, but otherwise arbitrarily large) level $k$, is it possible to find a value for $n$ so that some non-principal embedding of spin-$n$ gravity with Lobachevsky boundary conditions is dual to a unitary field theory?

Generalization to supersymmetric cases, along the lines of \cite{Henneaux:2012ny,Hanaki:2012yf}, a better understanding of the $n\to\infty$ limit (like hs$[\lambda]$ theory), and generalizations to models with local degrees of freedom like higher spin topologically massive gravity \cite{Chen:2011vp} would also be of interest.


Finally, there are plenty of other non-AdS boundary conditions available, some of which were presented in \cite{Gary:2012ms}, but not at the level of the present work: asymptotically Schr\"odinger, Lifshitz and warped AdS spacetimes. Some of these backgrounds allow for black hole solutions, and the study of black holes in higher spin gravity is of considerable interest, see \cite{Ammon:2012wc} and references therein. Including black holes requires an extension of the bulk-plus-boundary action \eqref{eq:nads1}-\eqref{eq:angelinajolie} to further boundary and corner terms \cite{Banados:2012ue}.
Apart from these less symmetric backgrounds, also the maximally symmetric cases of asymptotically dS \cite{Anninos:2011ui} or asymptotically flat \cite{Barnich:2006av} boundary conditions provide interesting fields for future research.

\acknowledgments

We thank Ralph Blumenhagen, Andrea Campoleoni, Alejandra Castro, Niklas Johansson, Sasha Polyakov and Per Sundell for discussions.

HA, MG and DG were supported by the START project Y435-N16 of the Austrian Science Fund (FWF). 
Additionally, HA was supported by the FWF project I952-N16 and DG by the FWF project P21927-N16.
RR was supported by the FWF project P-22000 N-16 and the project DO 02-257 of the Bulgarian National Science Foundation (NSFB).
This research was supported in part by the National Science Foundation under Grant No. NSF PHY11-25915.

\appendix

\section{Review of canonical analysis}\label{app:A}

In order to proceed with the canonical analysis it is convenient to use a $2+1$ decomposition of the action \eqref{eq:nads1} \cite{Banados:1994tn,Blagojevic:2002du}. 
	\begin{equation}\label{Intro:2+1}
		I_{\textrm{\tiny CS}}[A]=\frac{k}{4\pi}\,\int_{\mathbb{R}}\extd t\int_{\mathcal{D}}\extd^2x\,\epsilon^{ij}g_{ab}\left(\dot{A}^a_iA^b_j+A^a_0F^b_{ij}\right)\,,
	\end{equation}
with $F^a_{ij}=\partial_iA^a_j-\partial_jA^a_i+f^a{}_{bc}A^b_iA^c_j$, $A=A^a T_a$, $g_{ab}=\Tr(T_a T_b)$, $[T_a,\,T_b]=f^c{}_{ab} T_c$, $\epsilon^{ij}=\epsilon^{tij}$, dot denotes $\partial_t$, and we dropped boundary terms.
Calculating the canonical momenta $\pi_a^\mu\equiv\frac{\partial\mathcal{L}}{\partial\dot{A}^a_\mu}$ corresponding to the canonical variables $A^a_\mu$ generates primary constraints $\phi_a^\mu$.
	\begin{equation}
		\phi_a^0:=\pi_a^0\approx0\qquad\phi_a^i:=\pi_a^i-\frac{k}{4\pi}\,\epsilon^{ij}g_{ab}A^b_j\approx0
	\end{equation}
The Poisson bracket has its canonical form, $\{A^a_\mu(\textbf{x}),\pi_b^\nu(\textbf{y})\}=\delta^a_b\,\delta_\mu^\nu\,\delta^2(\textbf{x}-\textbf{y})$.
The canonical Hamiltonian density, up to boundary terms, is  given by
	\begin{equation}
		\mathcal{H}=-\frac{k}{4\pi}\,\epsilon^{ij}g_{ab} A^a_0F^b_{ij}\,.
	\end{equation}
The total Hamiltonian is then given as $\mathcal{H}_T=\mathcal{H}+u^a_\mu\phi_a^\mu$, where $u^a_\mu$ are Lagrange multipliers. 
Conservation of the primary constraints, $\dot{\phi}_a^\mu=\{\phi_a^\mu,\,\mathcal{H}_T\}\approx0$, leads to the following secondary constraints
	\begin{equation}
		\mathcal{K}_a\equiv-\frac{k}{4\pi}\,\epsilon^{ij}g_{ab}F^b_{ij}\approx0\,,\qquad
		D_iA^a_0-u^a_i\approx0\,,\label{Intro:Multiplier}
	\end{equation}
with the covariant derivative $D_iX^a=\partial_iX^a+f^a{}_{bc}A^b_iX^c$. 
Defining $\bar{\mathcal{K}}_a=\mathcal{K}_a-D_i\phi_a^i$ the total Hamiltonian can be expressed as a sum over constraints.
	\begin{equation}
		\mathcal{H}_T=A^a_0\,\bar{\mathcal{K}}_a+u^a_0\,\phi_a^0
	\end{equation}
The non-vanishing Poisson brackets between the constraints lead to the following algebra.
	\begin{subequations}
		\begin{align}
			\{\phi_a^i(\textbf{x}),\phi_b^j(\textbf{y})\}&=-\frac{k}{2\pi}\,\epsilon^{ij}g_{ab}\,\delta^2(\textbf{x}-\textbf{y})\\
			\{\phi_a^i(\textbf{x}),\bar{\mathcal{K}}_b(\textbf{y})\}&=-f_{ab}{}^c\phi_c^i\,\delta^2(\textbf{x}-\textbf{y})\\
			\{\bar{\mathcal{K}}_a(\textbf{x}),\bar{\mathcal{K}}_b(\textbf{y})\}&=-f_{ab}{}^c\bar{\mathcal{K}}_c\,\delta^2(\textbf{x}-\textbf{y})
		\end{align}
	\end{subequations}
Thus $\phi_a^0$ and $\bar{\mathcal{K}}_a$ are first class constraints and $\phi_a^i$ are second class constraints. The second class constraints are eliminated by introducing the Dirac bracket (denoted again by $\{,\}$), which turns out to be identical to the Poisson bracket, except for the relation $\{A^a_i(\textbf{x}),A^b_j(\textbf{y})\}=\frac{2\pi}{k}\,g^{ab}\epsilon_{ij}\,\delta^2(\textbf{x}-\textbf{y})$.

As next step we construct the canonical generators of gauge transformations using Castellani's algorithm. 
They are given by $G=\dot{\epsilon}(t)G_1+\epsilon(t)G_0$,
where the constraints $G_0$ and $G_1$ have to fulfill the relations $G_1=C_{\textrm{PFC}}$, $G_0+\{G_1,\mathcal{H}_T\}=C_{\textrm{PFC}}$, $\{G_0,\mathcal{H}_T\}=C_{\textrm{PFC}}$.
Here $C_{\textrm{PFC}}$ denotes a primary first class constraint. These relations are fulfilled for $G_1=\pi_a^0$ and  $G_0=\bar{\mathcal{K}}_a$. The smeared generator of gauge transformations then reads
	\begin{equation}
		\bar{\cal G}[\epsilon]=\int_\mathcal{D}\extd^2x\left(D_0\epsilon^a\pi_a^0+\epsilon^a\bar{\mathcal{K}}_a\right)\,.
\label{eq:app1}
	\end{equation}
The generator $\bar{\cal G}$ is not yet functionally differentiable.
\eq{
		\delta \bar{\cal G}[\epsilon]
				   = 
\textrm{regular} - \int_\mathcal{D}\extd^2x\,\partial_i\Big(\frac{k}{4\pi}\,\epsilon^{ij}g_{ab}\epsilon^a\,\delta A^b_j+\epsilon^a\,\delta\pi_a^i\Big)
}{Intro:DeltaG}
The first term is the bulk variation of the generator \eqref{eq:app1}.
The second term is a boundary term and spoils functional differentiability. In order to fix this one adds a suitable boundary term $Q$ to the canonical generator \eqref{eq:app1} such that the variation of this additional boundary term cancels exactly the boundary term in \eqref{Intro:DeltaG}.
	\begin{equation}
		\delta{\cal G}[\epsilon]=\delta \bar{\cal G}[\epsilon]+\delta Q[\epsilon]
\label{eq:app2}
	\end{equation}
with
	\begin{equation}
		\delta Q[\epsilon]=\int_\mathcal{D}\extd^2x\,\partial_i\Big(\frac{k}{4\pi}\,\epsilon^{ij}g_{ab}\epsilon^a\delta A^b_j+\epsilon^a\delta\pi_a^i\Big)\,.
	\end{equation}
Using the Stokes theorem and the fact that in the reduced phase space the constraint $\phi_a^i$ strongly equals to zero, the variation of the boundary charge simplifies to
	\begin{equation}
		\delta Q[\epsilon]=\frac{k}{2\pi}\oint_{\partial\mathcal{D}}\!\!\extd\varphi\, g_{ab}\epsilon^a\delta A^b_\varphi\,.
	\end{equation}

\section{Non-principal embedding of $sl(2)$ into $sl(3)$}\label{app:B}

For explicit calculations in the non-principal embedding of $sl(2)$ into $sl(3)$ we use the generators
        \begin{equation}
                L_0=\frac{1}{2}\left(
                        \begin{array}{ccc}
                                1&0&0\\
                                0&0&0\\
                                0&0&-1
                        \end{array}\right)\qquad
                L_1=\left(
                        \begin{array}{ccc}
                                0&0&0\\
                                0&0&0\\
                                1&0&0
                        \end{array}\right)\qquad
                L_{-1}=\left(
                        \begin{array}{ccc}
                                0&0&-1\\
                                0&0&0\\
                                0&0&0
                        \end{array}\right)
        \end{equation}
In addition to the $sl(2)$ generators there are two doublets
        \begin{align}
                \psi_{\frac{1}{2}}^{+}&=\left(
                        \begin{array}{ccc}
                                0&0&0\\
                                -1&0&0\\
                                0&0&0
                        \end{array}\right) &
                \psi_{-\frac{1}{2}}^{+}&=\left(
                        \begin{array}{ccc}
                                0&0&0\\
                                0&0&1\\
                                0&0&0   
                        \end{array}\right) \\      
                \psi_{\frac{1}{2}}^{-}&=\left(
                        \begin{array}{ccc}
                                0&0&0\\
                                0&0&0\\
                                0&1&0
                        \end{array}\right) &
                \psi_{-\frac{1}{2}}^{-}&=\left(
                        \begin{array}{ccc}
                                0&1&0\\
                                0&0&0\\
                                0&0&0   
                        \end{array}\right)      
        \end{align}
and one singlet
        \begin{equation}
                S=\frac{1}{3}\left(
                        \begin{array}{ccc}
                                -1&0&0\\
                                0&2&0\\
                                0&0&-1
                        \end{array}\right)\,.
        \end{equation}



\providecommand{\href}[2]{#2}\begingroup\raggedright\endgroup

\end{document}